\documentclass[secthm,dvips]{arxstspdf}
\usepackage{dcolumn,multirow}
\usepackage{graphicx}
\usepackage{flushend,stfloats}
\volume{24}
\issue{4}
\pubyear{2009}
\firstpage{503}
\lastpage{516}
\doi{10.1214/09-STS314}

\makeatletter
\newcolumntype{d}[1]{D{.}{.}{#1}}
\makeatother

\begin{document}
\begin{frontmatter}

\title{Robust Tests in Genome-Wide Scans under Incomplete Linkage Disequilibrium}
\runtitle{Robust Genome-Wide Scans}

\begin{aug}
\author[a]{\fnms{Gang} \snm{Zheng}\ead[label=e1,text=zhengg@nhlbi.nih.gov]{zhengg@nhlbi.nih.gov}\corref{}},
\author[a]{\fnms{Jungnam} \snm{Joo}\ead[label=e2,text=jooj@nhlbi.nih.\\ gov]{jooj@nhlbi.nih.gov}},
\author[b]{\fnms{Dmitri} \snm{Zaykin}\ead[label=e3]{zaykind@niehs.nih.gov}},
\author[a]{\fnms{Colin} \snm{Wu}\ead[label=e4,text=wuc@nhlbi.nih.gov]{wuc@nhlbi.nih.gov}}
\and
\author[a]{\fnms{Nancy} \snm{Geller}\ead[label=e5,text=gellern@nhlbi.nih.gov]{gellern@nhlbi.nih.gov}}
\runauthor{G. Zheng et al.}

\affiliation{National Heart, Lung and Blood Institute, National Heart,
Lung and Blood Institute,
National Institute of Environmental Health Sciences and National Heart,
Lung and Blood Institute}

\address[a]{Gang Zheng is Mathematical Statistician,
Jungnam Joo is Mathematical Statistician,
Colin Wu is Mathematical Statistician,
Nancy Geller is Director, Office of Biostatistics Research, National Heart, Lung, and Blood Institute,
6701 Rockledge Drive, MSC 7913, Bethesda, Maryland 20892, USA \printead{e1}; \printead*{e2,e4}; \printead*{e5}.}
\address[b]{Dmitri Zaykin is Biostatistics Branch, National Institute of Environmental
Health Sciences, Mail Drop A3-03, Research
Triangle Park, North Carolina 27709, USA \printead{e3}.}
\end{aug}

%
\begin{abstract}
Under complete linkage disequilibrium (LD), robust tests often have
greater power than Pearson's chi-square test and trend tests for the
analysis of case-control genetic association studies. Robust statistics
have been used in candidate-gene and genome-wide association studies
(GWAS) when the genetic model is unknown. We consider here a more
general incomplete LD model, and examine the impact of penetrances at
the marker locus when the genetic models are defined at the disease
locus. Robust statistics are then reviewed and their efficiency and
robustness are compared through simulations in GWAS of 300,000 markers
under the incomplete LD model. Applications of several robust tests to
the Wellcome Trust Case-Control
Consortium [\textit{Nature} \textbf{447} (2007) 661--678] are presented.
\end{abstract}


\end{frontmatter}
%

\section{Introduction}\label{sec1}

Genome-wide association studies (GWAS) have been used to detect true
associations between 100,000 to 500,000 genetic markers
(single-nucleotide poly\-morphisms---SNPs) and common or complex diseases
(e.g., Klein et al., \citeyear{klein}; Sladek et al., \citeyear{sladek}; WTCCC, \citeyear{WTCCC}).
Currently, up to a million SNPs are used in GWAS. A simple and initial
analysis of GWAS is a genome-wide scan, in which a statistical test is
applied to detect association one SNP at a time. Test statistics and/or
their $p$-values are obtained for all SNPs and ranked in order of their
statistical significance. After all SNPs are ranked, a prespecified
small proportion of SNPs from the top-ranked SNPs (or SNPs with
$p$-values less than a prespecified genome-wide threshold level) is
selected for further, more focused analyses, for example, haplotype
analysis, multi-marker analysis, fine mapping, imputation and
independent replication studies (see Hoh and Ott, \citeyear{hoh};
Marchini, Donnelly and Cardon, \citeyear{marchini}; Schaid et al., \citeyear{schaid}). The genome-wide scan has also been
shown to be cost-effective in two-stage designs for GWAS, in which
additional subjects are genotyped in the second stage for a small
portion of selected SNPs in the first stage (see Elston, Lin and Zheng, \citeyear{elston};
Thomas et al., \citeyear{thomas}). We focus on robust tests for GWAS in the single
stage designs.

Since only a small portion of top-ranked SNPs is selected in
genome-wide scans, it is important that the probability of at least one
SNP with true association being selected is high, for example, greater
than 80\% (Zaykin and Zhivotovsky, \citeyear{zaykin05}; Gail et al., \citeyear{gail}). The
probability that a SNP with true association is detected, confirmed and
replicated in later more focused analyses is often smaller. Hence, one
of the goals of genome-wide scans is to rank the SNPs with true
associations as near to the top as possible. Zaykin and Zhivotovsky
(\citeyear{zaykin05}) showed that the factors that mainly affect the rankings of true
SNPs include the total number of SNPs, the number of SNPs with true
associations, the genetic effects (genotype relative risks or odds
ratios), the sample size, power of the association test used, and
linkage disequilibrium (LD) between SNPs and the functional locus (the
true unknown disease locus). Most of the above factors are determined
by the study design, except the power of the test for association. The
common association tests include Pearson's chi-squared test (Pearson's
test, for short), the Cochran-Armitage trend tests (CATTs) and the
allelic test. Three CATTs are available depending on the underlying
genetic model (the mode of inheritance of the disease locus). Common
genetic models include recessive, additive, multiplicative and dominant
models. Overdominant and underdominant models may also be used, but
they are less common. The allelic test has performance similar to that
of the CATT under the additive model when the Hardy--Weinberg
equilibrium proportions hold (Sasieni, \citeyear{sasieni}; Guedj, Nuel and Prum, \citeyear{guedj}). Thus,
the allelic test is not considered here.

Intuitively, the most powerful test should be used in genome-wide
scans. For common and complex diseases, it is possible that there are
multiple functional loci with different genetic models, in particular,
for GWAS. The power of an association test depends on the underlying
genetic models of the functional loci, which, however, are unknown.
They could be any of the four common genetic models or none of them. In
addition, imperfect LD between functional and marker loci can modify
the underlying genetic model, further increasing uncertainty. In this
case, there is no uniformly most powerful test for a genome-wide scan.
It is known that the most efficient CATT is available when the genetic
model is known (Sasieni, \citeyear{sasieni}; Freidlin et al., \citeyear{freidlin02}). When the genetic
model is unknown, using a single CATT is not robust across a family of
genetics models. Therefore, in this situation, more robust tests have
been proposed for both candidate-gene studies and genome-wide scans
(Freidlin et al., \citeyear{freidlin02}; Sladek et al., \citeyear{sladek}; Zheng and Ng, \citeyear{zhengNg};
Gonzalez et al., \citeyear{gonzalez}; Joo et al., \citeyear{joo}). The performance of the robust
test statistics has been studied under the perfect LD model, that is,
the SNP is the same as the functional locus (see more discussion
later). This is, however, a strong assumption for GWAS. In particular,
when one of the models embedded into a robust test holds at the
functional locus, it remains unmodified at the marker locus. Therefore,
it is not surprising that robust tests based on the maximum of test
statistics over common genetic models often provide greater power than
Pearson's test and CATTs. However, when LD is imperfect, the induced
penetrance values at the marker are weighted averages of the causal
penetrances, where the weights are functions of LD. Thus, the imperfect
LD will change certain models, such as the dominant or the recessive
models, so that the heterozygote penetrance will have an intermediate
value between those for the homozygotes. Therefore, it is important to
investigate not only the exact form of such penetrance modifications,
but also its impact on the performance of the robust tests for association.

In this article we consider a general LD model with the standardized LD
parameter, $D'$ (Lewontin, \citeyear{lew64}), and study the properties of the
penetrances defined at the marker locus given the genetic model defined
at the functional locus. In addition to reviewing some common robust
tests for case-control association studies, we also compare their
performance under this general model with a varying $D'$. Using robust
tests when there is imperfect LD has not been studied perviously. The
perfect LD case, where the marker and the disease loci coincide, can be
obtained as a special case at $D'=1$, with an additional requirement of
equality of allele frequencies at the marker and the disease locus.
This implies a perfect correlation between the alleles at the two loci.
Under this general model, we also examine the effectiveness and
robustness of the genetic model selection procedure (Zheng and Ng,
\citeyear{zhengNg}). Simulation studies are conducted to compare the efficiency
robustness of various robust tests under this general model for
genome-wide scans of 300,000 SNPs. Applications of robust tests are
presented using real data from a GWAS (WTCCC, \citeyear{WTCCC}).

The rest of the article is organized as follows. In Section \ref{sec2} we
introduce notation, the case-control data and different genetic models.
The\break Hardy--Weinberg disequilibrium coefficient and its use to detect
the underlying genetic model is given in Section \ref{sec3}. Various robust
tests for candidate-gene analysis and GWAS will be reviewed under the
perfect LD model in Section \ref{sec4}. Section \ref{sec5} presents numerical results
based on the simulation studies. The performance of the model \mbox{selection}
procedure under the general LD model will be reported. Comparison of
several robust tests in analyzing genome-wide data is also presented.
Applications to real data are given in Section \ref{sec6}. Discussion and
conclusions are given in the final section.
\section{Genetic Models}\label{sec2}

\subsection{Notation and Data}\label{sec2.1}

Consider a case-control association study with $r$ cases and $s$
controls and a SNP with alleles $A$ and~$B$. Denote the population
frequencies of the alleles by $\operatorname{Pr}(B)=p$ and $\operatorname{Pr}(A)=p_c=1-p$. The three
genotypes of the SNP are denoted by $G_0=AA$, $G_1=AB$, and $G_2=BB$,
with the population frequencies $\operatorname{Pr}(G_i)=g_i$ for $i=0,1,2$. When the
Hardy--Weinberg equilibrium (HWE) proportions hold in the population,
$(g_0, g_1, g_2) = (p_c^2, 2pp_c, p^2)$. The case-control
data for the SNP can be displayed in a $2\times3$ contingency table
with the rows corresponding to case or control groups and the columns
to the three genotypes. The genotype counts for $(G_0,G_1,G_2)$ in
cases and controls are denoted by $(r_0,r_1,r_2)$ and $(s_0,s_1,s_2)$,
respectively. The genotype counts follow multinomial distributions:
$(r_0,r_1,r_2)\sim \operatorname{Mul}(r; p_0, p_1,\break p_2)$ and $(s_0,s_1,s_2)\sim \operatorname{Mul}(s;
q_0,q_1,q_2)$, where $p_i=\operatorname{Pr}(G_i|\mbox{case})$ and $q_i=\operatorname{Pr}(G_i|\mbox{control})$
for $i=0,1,2$. Under the null hypothesis of no association,
$H_0\dvtx p_i=q_i$ for all $i$.

Denote the penetrance of the SNP by $f_i=\operatorname{Pr}(\mbox{case}|\break G_i)$, and the
disease prevalence by $k=\operatorname{Pr}(\mbox{case})$. Then $p_i=g_i f_i/k$ and
$q_i=g_i(1-f_i)/(1-k)$. Hence, the null hypothesis becomes $H_0\dvtx
f_0=f_1=f_2=k$. For simplicity, we assume in this section there is only
one functional locus. Therefore, there is only one genetic model.

\subsection{Perfect LD Model}\label{sec2.2}

Under this model, the SNP is also the functional locus with equal
allele frequencies. The penetrances $f_i$, $i=0,1,2,$ defined earlier are
also penetrances of the function{al} locus. Genotype relative risks
(GRRs) are defined by $\lambda_i=f_i/f_0$ for $i=1,2,$ where $f_0$ is
the reference penetrance. Under the alternative hypothesis, allele $B$
is the risk allele if the probability of having the disease increases
with the number of $B$ alleles in the genotype. That is, $f_2\ge f_1\ge
f_0$ and $f_2>f_0$. These two constraints define a family of
constrained genetic models, which contains four commonly used genetic models:
%
\begin{equation}
\Lambda=\{(\lambda_1,\lambda_2)\dvtx \lambda_2\ge\lambda_1 \mbox{ and }
\lambda_2>1\}.\label{const1}
\end{equation}
We refer to $\Lambda$ as the constrained space for genetic models when
the risk allele is known. The null hypothesis corresponds to $H_0\dvtx
\lambda_1=\lambda_2=1$. The genetic model is recessive if $\lambda
_1=1$, additive if $\lambda_1=(1+\lambda_2)/2$, multiplicative if
$\lambda_1=\lambda_2^{1/2}$, and dominant if $\lambda_1=\lambda_2$. Let
$\lambda_2=\lambda$ for some $\lambda\ge1$. Then $\lambda_1$ can be
calculated using~$\lambda$ value under one of the four genetic models.
The first three letters of each model are used to indicate the genetic
model in the following, for example, REC stands for the recessive model.

Note that $\Lambda$ does not contain overdominant or underdominant
models, which occurs when $\lambda_1\ge\lambda_2\ge1, \lambda_1 > 1$
and $\lambda_2\ge1 \ge\lambda_1, \lambda_2>\lambda_1$, respectively.
These two models are less common compared to the other four genetic
models reviewed here.

\subsection{Incomplete LD Model}\label{sec2.3}

Under this model, the SNP of interest is not the functional locus.
Suppose the functional locus also has two alleles, denoted by $a$ and
$b$, with the population frequencies $\operatorname{Pr}(b)=q$ and $\operatorname{Pr}(a)=q_c=1-q$.
Assume that the SNP with alleles $A$ and $B$ is associated with the
disease through LD with the function{al} locus with alleles $a$ and
$b$. Table \ref{LD} represents the joint probabilities of the two loci,
in which $D=\operatorname{Pr}(Aa)-\operatorname{Pr}(A)\operatorname{Pr}(a)$ measures LD between the SNP and the
functional locus. When $D=0$, they are in linkage equilibrium. An
association between the SNP and a disease can be established when $|D|
> 0$ and when the two loci are linked.

\begin{table}[b]
\caption{Joint probabilities of the marker and function{al} locus under
incomplete LD model\label{LD}}
\begin{tabular*}{\columnwidth}{@{\extracolsep{\fill}}lccc@{}}
\hline
&\multicolumn{2}{c}{\textbf{Functional locus}} &\\
\cline{2-3} \\[-6pt]
\textbf{Marker} & $\bolds{a}$ & $\bolds{b}$ &\\
\hline
$A$ & $p_cq_c+D$ & $p_cq-D$ & $p_c$\\
$B$ & $ pq_c-D$ & $pq+D$ & $p$\\[6pt]
&$q_c$&$q$&1\\
\hline
\end{tabular*}
\end{table}

\begin{table*}[b]
\caption{\label{transp} Conditional probabilities in the transition
matrix (\protect\ref{trans})}
\begin{tabular*}{\textwidth}{@{\extracolsep{\fill}}ll@{}}
\hline
$\bolds{\operatorname{Pr}(G^*_i|G_j)}$&\textbf{Formula}\\
\hline
$\operatorname{Pr}(G^*_0|G_0)=\operatorname{Pr}(aa|AA)$ & $p^2_{Aa}/
(p^2_{Aa}+2p_{Aa}p_{Ab}+p^2_{Ab})=F_1^2$\\
$\operatorname{Pr}(G^*_0|G_1)=\operatorname{Pr}(aa|AB)$ & $p_{Aa}p_{Ba}/
(p_{Aa}p_{Ba}+p_{Aa}p_{Bb}+p_{Ab}p_{Ba}+p_{Ab}p_{Bb})=F_1F_2$\\
$\operatorname{Pr}(G^*_0|G_2)=\operatorname{Pr}(aa|BB)$ & $p^2_{Ba}/
(p^2_{Ba}+2p_{Ba}p_{Bb}+p^2_{Bb})=F_2^2$\\
$\operatorname{Pr}(G^*_1|G_0)=\operatorname{Pr}(ab|AA)$ & $2p_{Aa}p_{Ab}/
(p^2_{Aa}+2p_{Aa}p_{Ab}+p^2_{Ab})=2F_1F_3$\\
$\operatorname{Pr}(G^*_1|G_1)=\operatorname{Pr}(ab|AB)$ & $(p_{Aa}p_{Bb}+p_{Ab}p_{Ba})/
(p_{Aa}p_{Ba}+p_{Aa}p_{Bb}+p_{Ab}p_{Ba}+p_{Ab}p_{Bb})=F_1F_4+F_2F_3$\\
$\operatorname{Pr}(G^*_1|G_2)=\operatorname{Pr}(ab|BB)$ & $2p_{Ba}p_{Bb}/
(p^2_{Ba}+2p_{Ba}p_{Bb}+p^2_{Bb})=2F_2F_4$\\
$\operatorname{Pr}(G^*_2|G_0)=\operatorname{Pr}(bb|AA)$ & $p^2_{Ab}/
(p^2_{Ab}+2p_{Aa}p_{Ab}+p^2_{Ab})=F_3^2$\\
$\operatorname{Pr}(G^*_2|G_1)=\operatorname{Pr}(bb|AB)$ & $p_{Ab}p_{Bb}/
(p_{Aa}p_{Ba}+p_{Aa}p_{Bb}+p_{Ab}p_{Ba}+p_{Ab}p_{Bb})=F_3F_4$\\
$\operatorname{Pr}(G^*_2|G_2)=\operatorname{Pr}(bb|BB)$ & $p^2_{Bb}/
(p^2_{Ba}+2p_{Ba}p_{Bb}+p^2_{Bb})=F^2_4$\\
\hline
\end{tabular*}
\tabnotetext[]{}{$F_1=(p_cq_c+D)/p_c$, $F_2=(pq_c-D)/p$,
$F_3=(p_cq-D)/p_c$, $F_4=(pq+D)/p$}
\end{table*}

There are two commonly used measures of the relationship between the
SNP and the functional locus: $D'$ and the correlation between the
alleles $A$ and $a$. Denote $p_{Aa}=\operatorname{Pr}(Aa)$, $p_{Ab}=\operatorname{Pr}(Ab)$,
$p_{Ba}=\operatorname{Pr}(Ba)$, and $p_{Bb}=\operatorname{Pr}(Bb)$. Then
$D=p_{Aa}p_{Bb}-\break p_{Ab}p_{Ba}$. The measure $D'\in[-1,1]$ of Lewontin
(\citeyear{lew64}) is defined as
\begin{eqnarray*}
D'&=&\frac{D}{\min(q_cp, p_cq)}, \quad \mbox{if $D>0$};\\
&=&\frac{D}{\min(q_cp_c, pq)},\quad  \mbox{if $D\le0$}.
\end{eqnarray*}
When the SNP is identical to the functional locus (i.e., $A\equiv a$,
$B\equiv b$ and $p\equiv q$), $p_{Bb}=p$, $p_{Aa}=p_c$, and
$p_{Ab}=p_{Ba}=0$. Thus, $D'=1$. However, $D'=1$ can be reached when
the SNP is not identical to the functional locus (e.g., when $p \ne
q$). The correlation between the two alleles is defined as (Weir, \citeyear{weir})
\[
\operatorname{Corr}(A,a) = \frac{p_{Aa}p_{Bb}-p_{Ab}p_{Ba}}{\sqrt{p p_c q q_c}}.
\]
Note that the correlation reaches its maximum value only when $p = q$.
The LD model is \textit{complete} if $|D'|=1$ and \textit{perfect}
if $|\operatorname{Corr}(A,a)|=1$. In this article we assume the two loci have the same
allele frequencies. Thus, $D'$ and the correlation are equivalent. That
is, in this article the (im)perfect LD model is equivalent to the
(in)complete LD model.

In the simulations we specify $D'$, $p$ and $q$. Then, $D$ can be
calculated. Using Table \ref{LD}, the four haplotype frequencies
$p_{Aa}$, $p_{Ab}$, $p_{Ba}$ and $p_{Bb}$ can be obtained by replacing
$D$ in Table \ref{LD} by $D'\min(q_cp, p_cq)$ when $D\ge0$ (a similar
term is used when $D<0$).

The definition of a genetic model under the imperfect LD model differs
from that under the perfect LD model. Denote the genotypes at the
functional locus by $G^*_0=aa$, $G^*_1=ab$ and $G^*_2=bb$. The
penetrance of the function{al} locus is given by $f^*_i=\operatorname{Pr}(\mbox{case}|G^*_i)$ for $i=0,1,2$. Accordingly, define GRRs by $\lambda
^*_i=f^*_i/f^*_0$ for $i=1,2$. The penetrance of the SNP is the same as
before and still denoted by $f_i$. Denote $\mathbf{f}=(f_0,f_1,f_2)^t$,
$\mathbf{f}^*=(f^*_0,f^*_1,f^*_2)^t$, where $t$ is transpose and $\mathbf{P}^*=(\operatorname{Pr}(G^*_i|G_j))_{3\times3}$ and $\mathbf{P}=
(\operatorname{Pr}(G_i|G^*_j))_{3\times3}$ are $3\times3$ transition matrices.
Then we have
%
\begin{eqnarray}
\mathbf{f}&=&\mathbf{P}^{*t}\mathbf{f}^*\label{trans},\\
\mathbf{f}^*&=&\mathbf{P}^t\mathbf{f}\label{trans2}.
\end{eqnarray}
Under the perfect LD model, the two transition matrices are identity
matrices $\mathbf{P}^*=\mathbf{P}=\mathbf{I}$. The conditional probabilities in
(\ref{trans}) can be obtained using $\operatorname{Pr}(G^*_i|G_j)=\operatorname{Pr}(G^*_i, G_j)/\sum
_{l=0}^2 \operatorname{Pr}(G^*_l, G_j)$ under the Hardy--Weinberg proportions at both
SNP and functional locus, which are given in Table \ref{transp}. Note
that these are functions of the four haplotype frequencies. The
conditional probabilities in (\ref{trans2}) can be obtained similarly,
and can also be found in Nielsen and Weir (\citeyear{nielsen1}) and Hanson et al.
(\citeyear{hanson}), Table 3.

\begin{figure*}[b]

\includegraphics{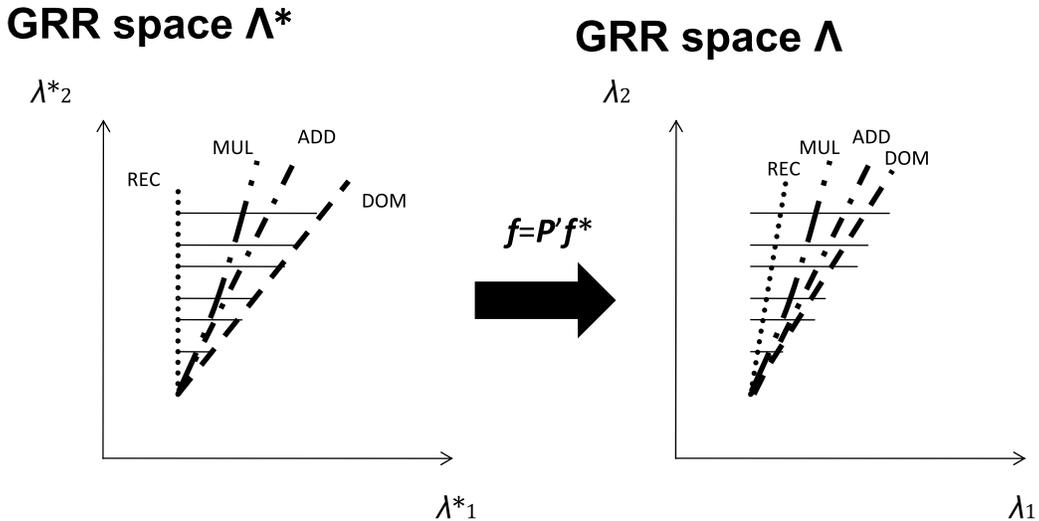}

\caption{Plots of the GRR spaces $\Lambda^*$ and $\lambda$ under the
inperfect LD model.}\label{fig1}
\end{figure*}

\subsection{Properties of Genetic Models under the Imperfect LD
Model}\label{sec2.4}

We defined genetic models using penetrances $(f_0,\break f_1,f_2)$ at the SNP
of interest. Under the imperfect LD model, the genetic model should be
defined at the functional locus using $(f^*_0,f^*_1,f^*_2)$. Thus, the
REC, ADD, MUL or DOM models correspond to $\lambda^*_1=1$, $\lambda
^*_1=(\lambda^*_2+1)/2$, $\lambda^*_1=\lambda^{*1/2}_2$, or $\lambda
^*_1=\lambda^*_2$, respectively. A constrained family of possible
genetic models at the functional locus is given by
%
\begin{equation}
\Lambda^*=\{(\lambda^*_1,\lambda^*_2)\dvtx \lambda^*_2\ge\lambda^*_1 \mbox
{ and } \lambda^*_2>1\}.\label{const2}
\end{equation}
Note that $\Lambda$ and $\Lambda^*$ are different under the imperfect
LD model, and they are linked by the two transition matrices in (\ref
{trans}) and (\ref{trans2}).
Under the imperfect LD model, applying Table \ref{transp} to $f_i=\sum
_{j=0}^2 \operatorname{Pr}(G^*_j|G_i)f^*_j$, we have
%
\begin{eqnarray}
\qquad f_0&=&f^*_0(F_1^2+2F_1F_3\lambda^*_1+F_3^2\lambda^*_2
),\label{f0}\\
f_1&=&f^*_0\{F_1F_2+(F_1F_4+F_2F_3)\lambda^*_1+F_3F_4\lambda^*_2\}
,\label{f1}\\
f_2&=&f^*_0(F_2^2+2F_2F_4\lambda^*_1+F_4^2\lambda^*_2).\label{f2}
\end{eqnarray}
The true disease model at the functional locus, defined using $(\lambda
^*_1,\lambda^*_2)$, is unknown. We study properties of the penetrances
$(f_0,f_1,f_2)$ or GRRs $(\lambda_1,\lambda_2)$ defined at the SNP
given $(\lambda^*_1,\lambda^*_2)$.

\begin{thm}\label{thm2.1} Under the imperfect LD model with $|D'| < 1$, if
$(\lambda^*_1,\lambda^*_2)\in\Lambda^*$ at the functional locus, then
$(\lambda_1,\lambda_2)\in\Lambda$ at the marker locus. Moreover, for
$(\lambda^*_1,\lambda^*_2)\in\Lambda^*-\{(1,1)\}$, if $\lambda^*_1=1$
(or $\lambda^*_1=\lambda^*_2$), then $\lambda_1>1$ (or $\lambda
_2>\lambda_1$).
\end{thm}

\begin{pf} Using $F_2-F_1=-D/(pp_c)=-(F_4-F_3)$ and (\ref{f0}) to
(\ref{f2}), we obtain
%
\begin{eqnarray}
f_1-f_0&=&\frac{f^*_0D}{pp_c}\{F_1(\lambda^*_1-1)+ F_3(\lambda
^*_2-\lambda^*_1)\},\label{e1}\\
f_2-f_1&=&\frac{f^*_0D}{pp_c}\{F_2(\lambda^*_1-1)+ F_4(\lambda
^*_2-\lambda^*_1)\}.\label{e2}
\end{eqnarray}
It follows that $f_2\ge f_1\ge f_0$ and $f_2>f_0$ when $f^*_2\ge
f^*_1\ge f^*_0$ and $f^*_2>f^*_0$. The proof of the second claim is
trivial using the above two expressions and that, from Table \ref{LD},
all $F_i$, $i=1,2,3,4$, are positive.
\end{pf}

Theorem \ref{thm2.1} shows that when the GRRs are constrained in $\Lambda^*$ at
the functional locus, they are also constrained to a subset of $\Lambda
$ at the SNP when $|D'| < 1$. In addition, when the true disease model
is either REC or DOM at the functional locus, it is no longer REC or
DOM at the SNP, respectively. They are ``closer'' to the ADD/MUL models.
The implication of this finding is that one will not see a pure DOM or
REC model at the marker locus if the constrained model space $\Lambda
^*$ is considered at the functional locus. It also provides a rationale
for the genetic model selection approach (Zheng and Ng, \citeyear{zhengNg}) in that
an ADD/MUL is always chosen unless there is strong evidence to indicate
the REC or DOM models.

Even though the REC (or DOM) model at the functional locus is no longer
retained at the SNP when $|D'| < 1$, the ADD (or MUL) model is
retained. Dividing (\ref{e1}) and (\ref{e2}) by $f_0$, we obtain
%
\begin{equation}
2\lambda_1-1-\lambda_2=\frac{f^{*}_0D^2}{f_0p^2p_c^2}(2\lambda
^*_1-1-\lambda^*_2).\label{add}
\end{equation}
Using (\ref{f0}) to (\ref{f2}) to expand $\lambda_2-\lambda
^2_1=(f_2f_0-f_1^2)/f^2_0$ and $(F_2F_3-F_1F_4)^2=D^2/(p^2p^2_c)$, we obtain
%
\begin{equation}
\lambda_2-\lambda^2_1=\frac{f^{*2}_0D^2}{f^2_0p^2p_c^2}(\lambda
^*_2-\lambda^{*2}_1).\label{mul}\\
\end{equation}
The above two equations lead directly to the following result.

\begin{thm}\label{thm2.2} Under the imperfect LD model with $|D'|<1$, when the
genetic model is ADD ($\lambda^*_1=(1+\lambda^*_2)/2$) or MUL ($\lambda
^*_2=\lambda^{*2}_1$) at the functional locus, the same model is
retained at the marker locus.
\end{thm}

Figure \ref{fig1} displays the mapping of genetic models from $\Lambda
^*$ to $\Lambda$ under the imperfect LD model. If we still define a
genetic model at the marker locus under the imperfect LD model, then,
using (\ref{trans2}) and a table similar to Table \ref{transp}, the REC
or DOM models at the marker locus would correspond to the underdominant
or overdominant models at the functional locus,
respectively.\looseness=1


\section{The Hardy--Weinberg Disequilibrium Coefficient and Genetic
Model Selection}\label{sec3}

The Hardy--Weinberg disequilibrium (HWD) coefficient in cases or
between cases and controls has been used to detect association (Nielsen, Ehm and Weir, \citeyear{nielsen};
Zaykin and Nielsen, \citeyear{zaykin00}; Song and Elston, \citeyear{song}). In
addition, it can also be used to detect the underlying genetic model at
the marker locus (Wittke-Thompson, Pluzhnikov and Cox, \citeyear{wittke}; Zheng and Ng, \citeyear{zheng08}).
In this section we first review the HWD coefficient and how it can be
used to detect the genetic model at the SNP of interest. Then we study
whether it can still be used to detect the genetic model which is
defined at the functional locus under the imperfect LD model.

Using the notation in Section \ref{sec1}, the HWD coefficient at the SNP with
alleles $A$ and $B$ is given by (Weir,~\citeyear{weir})
\begin{eqnarray*}
\Delta&=&\operatorname{Pr}(AA)-\{\operatorname{Pr}(AA)+\operatorname{Pr}(AB)/2\}^2\\
&=&g_2-(g_2+g_1/2)^2.
\end{eqnarray*}
In cases and controls, it is denoted by $\Delta_1$ and $\Delta_0$,
respectively, and given by
\begin{eqnarray*}
\Delta_1&=&p_2-(p_2+p_1/2)^2\quad  \mbox{and}\\
 \Delta_0&=&q_2-(q_2+q_1/2)^2.
\end{eqnarray*}
Substituting $p_i=g_i f_i/k$ and $q_i=g_i(1-f_i)/(1-k)$ under the
Hardy--Weinberg proportions ($\Delta=0$), one has (Wittke-Thompson, Pluzhnikov and Cox, \citeyear{wittke}; Zheng and Ng, \citeyear{zhengNg})
%
\begin{eqnarray}
\hspace*{25pt}\Delta_1&=&\frac{f^2_0p^2p^2_c}{k^2}(\lambda_2-\lambda^2_1
),\label{delta1}\\
\Delta_0&=&\frac{f^2_0p^2p^2_c}{(1-k)^2}(2\lambda_1-1-\lambda
_2-f_0\lambda^2_1+f_0\lambda_2).\label{delta0}
\end{eqnarray}
Using the signs of $(\Delta_1,\Delta_0)$, Zheng and Ng (\citeyear{zheng08}) divided
$\Lambda$ in (\ref{const1}) into four mutually exclusive regions $R_1$
to $R_4$. The signs in the four regions are $(\Delta_1,\Delta_0)=(+,-)$
in $R_1$, $(-,-)$ in $R_2$, $(-,-)$ in $R_3$, and $(-,+)$ in $R_4$. The
REC model belongs to $R_1$ and the DOM model belongs to $R_4$. The
region $R_2$ is bounded by the ADD and MUL models (see Figure 1 of Zheng
and Ng, \citeyear{zhengNg}). Therefore, under the REC model (defined at the SNP with
$\lambda_1=1$), $\Delta_1>0$ and $\Delta_0<0$, and under the DOM model,
$\Delta_1<0$ and $\Delta_0>0$. Zheng and Ng (\citeyear{zhengNg}) used $\partial\Delta
=\Delta_1-\Delta_0$ as a genetic model {indicator}. The REC model
implies that $\partial\Delta>0$, while the DOM model implies $\partial
\Delta<0$. A normalized test statistic based on $\widehat{\partial\Delta
}=\widehat{\Delta}_1-\widehat{\Delta}_0$, where $\widehat{p}_i=r_i/r$
and $\widehat{q}_i=s_i/s$, is given
\begin{eqnarray*}
Z_{\mathrm{HWDTT}}&=&\frac{(rs/n)^{{1}/{2}}\widehat{\partial\Delta
}}{\{1-n_2/n-n_1/(2n)\}\{n_2/n+n_1/(2n)\}}\\
&\sim& N(0,1)
\end{eqnarray*}
under $H_0$ and referred to as the HWD trend test (HWDTT) (Song and
Elston, \citeyear{song}). It is used to select a genetic model (Zheng and Ng,
\citeyear{zhengNg}). Given that~$B$ is the risk allele, the ADD (or MUL) model is
chosen unless there is strong evidence to indicate a REC model or a DOM
model. When $Z_{\mathrm{HWDTT}}>1.645$, the REC model is selected; when
$Z_{\mathrm{HWDTT}}<-1.645$, the DOM model is selected.

Under the imperfect LD model, using (\ref{mul}) and (\ref{add}), (\ref
{delta1}) and (\ref{delta0}) can be written as
\begin{eqnarray*}
\Delta_1&=&\frac{f^{*2}_0D^2}{k^2}(\lambda^*_2-\lambda^{*2}_1
),\\
\Delta_0&=&\frac{f_0f^{*}_0D^2}{(1-k)^2}(2\lambda^*_1-1-\lambda
^*_2-f^*_0\lambda^{*2}_1+f^*_0\lambda^*_2).
\end{eqnarray*}
Comparing the above $(\Delta_1,\Delta_0)$ with (\ref{delta1}) and (\ref
{delta0}), we see that the signs of $(\Delta_1,\Delta_0)$ do not change
when the genetic model is defined at the functional locus. Hence, the
model selection procedure of Zheng and Ng (\citeyear{zhengNg}) can still be used.

\section{Robust Tests}\label{sec4}

\subsection{Pearson's Test and CATTs}\label{sec4.1}

Given the case-control data for a single SNP, $(r_0,\break r_1,r_2)$ and
$(s_0,s_1,s_2)$, denote $n_i=r_i+s_i$ for $i=0,1,2$ and
$n=n_0+n_1+n_2$. Pearson's test can be written as
\begin{eqnarray*}
T_{\chi^2}&=&\sum_{i=0}^2(r_i-n_ir/n)^2/(n_ir/n)\\
&&{}+\sum_{i=0}^2
(s_i-n_is/n)^2/(n_is/n),
\end{eqnarray*}
which asymptotically follows a chi-squared distribution with 2 degrees
of freedom (df) under $H_0$. The CATT with a score $x\in[0,1]$ is
given by
\begin{eqnarray*}
Z_{x}&=&n^{1/2} \Biggl(n\sum_{j=0}^2 x_jr_j - r \sum_{j=0}^2 x_j n_j\Biggr)\\
&&{}\Big/
[rs\{n(n_1+4n_2)-(n_1+2n_2)^2\}]^{1/2},
\end{eqnarray*}
where $(x_0,x_1,x_2)=(0,x,1)$. Under $H_0$, $Z_x$ asymptotically
follows the standard normal distribution $N(0,1)$ for a given $x$.
Optimal scores for REC,\break ADD/MUL and DOM models are $x=0,1/2$ and $1$.

When the genetic model is unknown, $Z_{1/2}$ is often used. There is a
trade-off between $T_{\chi^2}$ and $Z_x$ with $x=1/2$. Pearson's test
is more robust but less powerful, in particular, under the ADD or DOM
models, while the trend test is more powerful under the ADD or DOM
models but less robust when the score $x$ is misspecified. Pearson's
test is identical to the trend test $Z^2_x$ with
$x=(r_1/n_1-r_0/n_0)/(s_1/n_1-s_0/n_0)$ (Yamada and Okada, \citeyear{yamada};
Zheng, Joo and Yang, \citeyear{zheng09}). In practice, however, $x$ is prespecified. Thus, this
condition is rarely satisfied.

\subsection{MAX}\label{sec4.2}

To avoid the trade-off between Pearson's test and the CATT, one
approach is to consider maximum tests. A~typical maximum test is given
by (Freidlin et al., \citeyear{freidlin02}; Sladek et al., \citeyear{sladek})
\[
\mbox{MAX}_3=\max\{|Z_0|,|Z_{1/2}|,|Z_1|\}.
\]
Other versions of maximum tests are also used, for example, $\mbox
{MAX}=\sup_{x\in[0,1]}|Z_x|$ (Davies, \citeyear{davies77,davies87}), the maximum of
three likelihood ratio tests under various genetic models (Gonz\'{a}lez
et al., \citeyear{gonzalez}), and for a quantitative trait (Lettre, Lange and Hirschhorn, \citeyear{lettre}).

Computational aspects of maximum tests have been discussed by Conneely
and Boehnke (\citeyear{conneely}) and Li et al. (\citeyear{li08a}). The empirical distribution of
$\mbox{MAX}_3$ can be obtained from simulation using the joint
multivariate normal distribution of the CATTs considering asymptotic
null correlations among them (Freidlin et al., \citeyear{freidlin02}) or from a
parametric bootstrap procedure by generating data using
$(r_0,r_1,r_2)\sim\break \operatorname{Mul}(r; \widehat{p}_0,\widehat{p}_1, \widehat{p}_2)$
and $(s_0,s_1,s_2)\sim \operatorname{Mul}(s; \widehat{p}_0,\widehat{p}_1, \widehat
{p}_2)$, where $\widehat{p}_i=n_i/n$. A simpler algorithm to find the
asymptotic and empirical null distributions of $\mbox{MAX}_3$ is
recently proposed (Zang, Fung and Zheng, \citeyear{zang}). The asymptotic null distribution
of $\mbox{MAX}_3$ is a function of the minor allele frequency (MAF) of
the SNP. In a genome-wide scan to rank a large number of SNPs, Li et
al. (\citeyear{li08b}) demonstrated that ranking can be done easily by the values
of $\mbox{MAX}_3$ rather than by their $p$-values. Hence, there is no
need to calculate the $p$-values of $\mbox{MAX}_3$, even though the
$p$-values of $\mbox{MAX}_3$ are more comparable across SNPs.

\subsection{MIN2}\label{sec4.3}

An alternative approach used by WTCCC (\citeyear{WTCCC}) utilizes both Pearson's
test and the CATT $Z_{1/2}$. WTCCC (\citeyear{WTCCC}) proposed to use the minimum
of the $p$-values of $T_{\chi^2}$ and $Z_{1/2}$ to scan all the SNPs.
SNPs with the minimum $p$-value less than a threshold level were retained
for further analyses. Joo et al. (\citeyear{joo}) denoted the minimum of the two
$p$-values by
\[
\mbox{MIN2}=\min\{p_{T_{\chi^2}}, p_{Z_{1/2}}\}
\]
and obtained its asymptotic null distribution and its $p$-value, denoted
by $p_{\mathrm{MIN2}}$. The key formula to find the distribution and
$p$-value for MIN2 is the joint distribution of Pearson's test and
$Z_{1/2}$ under $H_0$, which is given by (Joo et al., \citeyear{joo})
\begin{eqnarray*}
&&\operatorname{Pr}(Z^2_{1/2}<t_1, T_{\chi^2}<t_2)\\
&&\quad=1-\frac{1}{2}e^{-{t_1}/{2}}-{1}/{2}e^{-{t_2}{/2}}\\
&&\qquad{}+\frac{1}{2\pi}\int_{t_1}^{t_2}e^{-{v}/{2}}\arcsin\biggl(\frac
{2t_1}{v}-1\biggr)\,dv,
\end{eqnarray*}
when $t_1<t_2$, and $\operatorname{Pr}(Z^2_{1/2}<t_1, T_{\chi^2}<t_2) =
1-\break \exp(-t_2/2)$ when $t_1>t_2$. Unlike $\mbox{MAX}_3$, the asymptotic
null distribution of MIN2 does not depend on the MAFs of SNPs. Hence,
MIN2 itself can be used to rank all SNPs, which results in the same
ranks as when the $p$-value of MIN2 is used. Joo et al. (\citeyear{joo})
demonstrated that $p_{\mathrm{MIN2}}>\mathrm{MIN2}$, because
$Z^2_{1/2}$ and $T_{\chi^2}$ are correlated under the alternative
hypothesis. Thus, MIN2 itself cannot be used as the $p$-value.

\subsection{The Genetic Model Selection (GMS) Procedure}\label{sec4.4}

The GMS procedure is an adaptive approach. It contains two phases. In
phase 1 the underlying genetic model is detected using the value and
sign of $Z_{\mathrm{HWDTT}}$ (Song and Elston, \citeyear{song}; see also Section \ref{sec3}).
Once the model is selected (REC, ADD/MUL or DOM), in the second phase,
the CATT optimal for the selected model is applied to test for
association. For example, if the REC model is selected using the HWDTT,
$Z_0$ would be used in phase 2 to test for association. Since the
analyses in the two phases are correlated, Zheng and Ng (\citeyear{zheng08}) derived
the asymptotic null correlation for the GMS. This correlation is
incorporated in the distribution of the test statistics to control for
the Type I error. Like MIN2, computing the $p$-value of the GMS requires
integrations. Like $\mbox{MAX}_3$, the GMS can be used to rank SNPs
(Zheng et al., \citeyear{zheng09b}). Using test statistics to directly rank SNPs is
easier than using $p$-values of the GMS. Since the GMS depends on which
allele is the risk allele or whether the minor allele is the risk
allele, for each SNP, we first determine the risk allele ($B$ is risk
allele if $Z_{1/2}>0$). If the risk allele is $B$, then the above GMS
can be applied. Otherwise, we can switch the two alleles and apply the
above GMS.

\subsection{Other Tests}\label{sec4.5}

Balding (\citeyear{balding}) provided an excellent review of statistical methods for
the analysis of association studies. Two other robust two-phase tests
are also available that we do not include here. One feature of these
methods is that the test statistics in two phases are asymptotically
independent under $H_0$\break (Zheng, Song and Elston, \citeyear{zheng07}, Zheng et al., \citeyear{zheng08}). In this case, the
second phase can be used as a ``self-replication,'' an idea proposed in
van Steen et al. (\citeyear{van}). Alternatively, the significance level $\alpha$
can be decomposed to $(\alpha_1,\alpha_2)$ such that $\alpha_1\alpha
_2=\alpha$, where $\alpha_1$ is used for the phase 1 analysis and
$\alpha_2$ for the phase 2 analysis. The null hypothesis is rejected
when analyses in both phases are significant at their corresponding
levels. Choices of $\alpha_1$ and $\alpha_2$ with $\alpha_1\alpha=\alpha
$ in GWAS were discussed in Zheng, Song and Elston (\citeyear{zheng07}), Zheng et al. (\citeyear{zheng08}). Another robust
test is the constrained likelihood ratio test (LRT) (Wang and
Sheffield, \citeyear{wang}). It is similar to the LRT except that the alternative
space is restricted to $\Lambda-\{(1,1)\}$. The performance of the
constrained LRT is similar to that of $\mathrm{MAX}_3$ described above.
Thus, we only consider $\mathrm{MAX}_3$ here.

\subsection{Why Robust Tests?}\label{sec4.6}

One of the reasons that we use robust tests in GWAS is that there might
be multiple functional loci for a given disease. The modes of
inheritance or genetic models may differ from one functional locus to
the other. Another reason for using robust tests is the distortion of
the actual genetic model at the marker locus due to incomplete LD,
which further amplifies uncertainty about the model. Thus, robust tests
are generally preferred. We use efficiency robustness to measure
robustness (Gastwirth, \citeyear{gast}). A test $T_1$ is said to have greater
efficiency robustness than a test $T_2$ if the worst asymptotic
relative efficiency of $T_1$ to the asymptotically optimal test across
all genetic models is higher than the worst asymptotic relative
efficiency of $T_2$. The CATT $Z_{1/2}$ optimal for the ADD model is
most robust among all trend tests when the genetic models are
constrained in $\Lambda$. Pearson's test is also robust because it does
not require the genetic models to be constrained or the alternative
hypothesis to be ordered. When restricting to $\Lambda$, tests more
robust than $Z_{1/2}$ are available. $\mbox{MAX}_3$ and GMS are two
examples. They both have greater efficiency robustness than Pearson's
test and $Z_{1/2}$ (Freidlin et al., \citeyear{freidlin02}; Zheng and Ng, \citeyear{zheng08}). On the
other hand, combining information of both Pearson's test and $Z_{1/2}$,
MIN2 is also more efficiency robust than either Pearson's test or
$Z_{1/2}$. Three robust tests, $\mbox{MAX}_3$, GMS and MIN2, appear to
have comparable efficiency robustness in candidate-gene studies (Joo et
al., \citeyear{joo}).

In genome-wide scans it is desirable to locate the SNPs representing
true association as near the top as possible, where all SNPs compete
for the top ranks. Under the complete LD model, Zheng et al. (\citeyear{zheng09b})
conducted simulation studies comparing the three robust methods in
ranking 300,000 SNPs,\break among which there were 6 functional loci with
different genetic models, MAFs and GRRs (from 1.25 to 1.5). The results
showed that the GMS slightly outperforms MIN2 and $\mbox{MAX}_3$ when
the top 5000 SNPs were selected. The criteria used for comparison
included the probability that the top 5000 SNPs contained at least one
SNP with true association, as well as the minimum and average ranks of
SNPs with true associations among the top 5000 SNPs. We will conduct
similar simulation studies in Section \ref{sec5} under the inperfect LD model.
The reason that we choose the top 5000 SNPs rather than a smaller
number, say, the top 100, is that the SNPs with true association are
not always ranked near the top, especially for a small GRR between 1.2
and 1.5 and small sample sizes (Zaykin and Zhivotovsky, \citeyear{zaykin05}). If we
examine the top 100 list with 250 cases and 250 controls (the sample
sizes that we used in our simulation studies), the probability that the
list of the top 100 SNPs contains a true association is less than 0.50.

\section{Simulation Studies}\label{sec5}

\subsection{The GMS Procedure under the Imperfect LD Model}\label{sec5.1}

We first conducted simulation studies to estimate the distribution of
genetic models selected by the GMS. We chose disease prevalence $k=0.1$
and GRR $\lambda^*_2=2$ at the functional locus. Then $\lambda^*_1$ was
obtained using $\lambda^*_2$ and a given genetic model at the
functional locus. We considered 0.1, 0.3 and 0.5 for the equal MAFs at
a SNP ($p$) and a functional locus ($q$). This allows us to compare the
frequencies of the different models selected when $D'=1.0$, 0.8 and
0.6. With equal allele frequencies $p=q$, Corr$(A,a) = D'$. In each of
10,000 replicates, 250 cases and 250 controls were simulated from
multinomial distributions in which the penetrances at a SNP were
calculated using (\ref{f0}) to (\ref{f2}). When the GMS did not select
REC or DOM, the ADD or MUL models are used and denoted here by A/M.
Results are reported in Table \ref{GMS}.

\begin{table*}
\caption{Distributions of genetic models selected by the
GMS using the HWDTT (\%): Disease prevalence $k=0.1$, the GRR at the
functional locus $\lambda^*_2=2$ with 250 cases and 250 controls and
10,000 replicates}\label{GMS}
\begin{tabular*}{\textwidth}{@{\extracolsep{\fill}}lcd{2.1}d{2.1}d{2.1}d{2.1}d{2.1}d{2.1}d{2.1}d{2.1}d{2.1}@{}}
\hline
& & \multicolumn{9}{c@{}}{$\bolds{D'}$\textbf{/selected models (A/M
$\bolds{=}$
ADD/MUL)}}\\
\ccline{3-11}\\[-6pt]
&&\multicolumn{3}{c}{\textbf{1.0}}&\multicolumn{3}{c}{\textbf{0.8}}&\multicolumn{3}{c@{}}{\textbf{0.6}}\\
\ccline{3-5,6-8,9-11}\\[-6pt]
\multicolumn{1}{@{}l}{\multirow{2}{30pt}[8pt]{\textbf{MAF} $\bolds{p=q}$}}
&\multicolumn{1}{c}{\multirow{2}{30pt}[8pt]{\centering\textbf{True model}}} & \multicolumn{1}{c}{\textbf{REC}} & \multicolumn{1}{c}{\textbf{A/M}} & \multicolumn{1}{c}{\textbf{DOM}} & \multicolumn{1}{c}{\textbf{REC}}
& \multicolumn{1}{c}{\textbf{A/M}} & \multicolumn{1}{c}{\textbf{DOM}} & \multicolumn{1}{c}{\textbf{REC}} &
\multicolumn{1}{c}{\textbf{A/M}} & \multicolumn{1}{c@{}}{\textbf{DOM}} \\
\hline
0.1 &REC&23.3&76.3& 0.4&14.6&84.4& 1.0& 3.0&90.6& 6.4\\
&ADD& 2.6&88.9& 8.5& 2.4&90.2& 7.4& 2.9&90.8& 6.3\\
&MUL& 3.4&90.3& 6.3& 3.7&91.1& 5.2& 3.8&90.8& 5.4\\
&DOM& 0.1&60.1&39.8& 0.3&76.1&23.6& 1.0&84.8&14.2\\
0.3 &REC&67.5&32.5& 0.0&39.4&60.4& 0.2&18.6&80.6& 0.8\\
&ADD& 2.2&88.9& 8.9& 3.1&89.3& 7.6& 3.7&89.6& 6.7\\
&MUL& 4.8&90.7& 4.5& 5.0&90.4& 4.6& 5.2&90.1& 4.7\\
&DOM& 0.0&32.8&67.2& 0.1&61.4&38.5& 0.7&80.2&19.1\\
0.5 &REC&66.0&34.0& 0.0&36.8&63.1& 0.2&18.3&80.9& 0.8\\
&ADD& 2.6&89.0& 8.4& 3.3&89.6& 7.1& 3.7&90.8& 5.5\\
&MUL& 5.4&89.9& 4.7& 5.0&90.1& 4.9& 5.2&89.9& 4.9\\
&DOM& 0.0&36.2&63.8& 0.1&63.9&36.0& 0.8&81.2&18.0\\
\hline
\end{tabular*}
\end{table*}

When the true model is REC or DOM at the functional locus, the
frequencies that the model selected by the GMS at the marker locus is
REC or DOM decreases dramatically when $D'$ becomes small. For example,
when $p=q=0.3$, the frequency of selecting REC at the marker locus is
about 67.5\% when the true model at the functional locus is REC, and
$D'=1$. This frequency declines to 18.6\% when $D'=0.6$. These
frequencies, however, are not sensitive when the true model at the
functional locus is either ADD or MUL. The findings are consistent with
Theorems \ref{thm2.1} and \ref{thm2.2}. Given the genetic model space $\Lambda^*$ at the
functional locus, the genetic model space at the marker locus $\Lambda$
is shifted toward the center of the space $\Lambda^*$ corresponding to
the ADD/MUL models.

\begin{table}[b]
\caption{GRRs $(\lambda_1,\lambda_2)$ at a SNP given GRR
$\lambda^*_2=2$ at the functional locus: $p=q=0.3$. When $D'=1$,
$\lambda^*_i=\lambda_i$ for $i=1,2$}\label{lambda}
\begin{tabular*}{\columnwidth}{@{\extracolsep{\fill}}lccc@{}}
\hline
&\multicolumn{3}{c@{}}{$\bolds{D'/(\lambda_1,\lambda_2)}$}\\
\ccline{2-4}\\[-6pt]
\multicolumn{1}{@{}l}{\multirow{2}{30pt}[8pt]{\textbf{True model}}} & \textbf{1.0} & \textbf{0.8} & \textbf{0.6}\\
\hline
REC &(1.00, 2.00) & (1.05, 1.73) & (1.07, 1.50)\\
ADD &(1.50, 2.00) & (1.38, 1.75) & (1.27, 1.54) \\
MUL &(1.41, 2.00) & (1.22, 1.48) & (1.24, 1.53)\\
DOM &(2.00, 2.00) & (1.67, 1.77) & (1.43, 1.57)\\
\hline
\end{tabular*}
\end{table}

Table \ref{lambda} reported the GRRs at the marker locus given those at
the functional locus. Note that when the true model is ADD ($\lambda
^*_1=(1+\lambda^*_2)/2$) or MUL ($\lambda^{*2}_1=\lambda^*_2$), the
GRRs at the marker locus follow the same models. However, $\lambda_i$
are smaller than $\lambda^*_i$. Similar patterns are observed when the
true model is REC or DOM, except that $\lambda_1$ is slightly greater
than $\lambda^*_1$ under the REC model.

\subsection{Comparison of Robust Tests in GWAS under the Imperfect LD
Model}\label{sec5.2}

In Table \ref{GMS} when the true model is REC or DOM at the functional
locus, the GMS could not select REC or DOM at the marker locus. This,
however, does not mean that the GMS cannot improve power or chances of
true discoveries when $|\operatorname{Corr}(A,a)| < 1$. On the contrary, owing
to the shrinkage of the genetic model space and that the GMS only
selects a model at the marker locus, it can be viewed as selecting an
appropriately induced model at the marker locus. Our next simulation
will examine the performance of robust tests under the imperfect LD
model. The simulation procedure follows the one used in Zheng et al.
(\citeyear{zheng09b}). We simulated genotype counts for each of 300,000 SNPs, among
which 6 SNPs have true associations and $D'=0.8$ with MAF of 0.2 at the
functional loci. When $D'=1$, the number of functional loci is also 6.
However, when $D'=0.8$, we assume the number of functional loci equals
the number of different genetic models in the simulation. Zheng et al.
(\citeyear{zheng09b}) considered the perfect LD model that corresponds to $|D'|=1$ or
$|\operatorname{Corr}(A,a)| = 1$. Their results are repeated here for
comparison. The MAFs of 6 true SNPs from the genetic models listed in
the titles of Tables \ref{ranks} and \ref{ranks2} were 0.1821, 0.2943,
0.1078, 0.4459, 0.1620 and 0.1825. These are also given in Zheng et al.
(\citeyear{zheng09b}) and in Li et al. (\citeyear{li08b}). MAFs for the rest of the null SNPs
were simulated from a uniform distribution $U(0.1, 0.5)$. The GRRs for
the functional loci were all 1.25 (or 1.50). We applied five robust
tests ($Z_{1/2}$, Pearson's test $T_{\chi^2}$, GMS, MIN2 and $\mbox
{MAX}_3$) to rank all SNPs and the top 5000 SNPs were selected from
each of 200 replicates. The criteria to compare the performance of
robust tests include the probability (prob \%) of at least one true SNP
being selected among the top 5000 SNPs, the average number of true
SNPs among the top, and the mean of the minimum ranks of the true SNPs
among the top. The results are presented in Table \ref{ranks} (2 REC, 1
ADD, 1 MUL and 2 DOM SNPs) and Table \ref{ranks2} (1 REC, 2 ADD, 2 MUL
and 1 DOM SNPs).

First, when $D'=1$ (Zheng et al., \citeyear{zheng09b}), the GMS outperforms other
tests under all three criteria, while Pearson's test had the worst
performance. When $D'=0.8$, however, the GMS and $Z_{1/2}$ had similar
performances, which together outperform other tests using the three
criteria. This finding is consistent to our results in Theorems \ref{thm2.1} and
\ref{thm2.2} about the genetic models under the imperfect LD model.

\begin{table*}
\caption{\label{ranks}Genome-wide scans of 300,000 SNPs containing 6
true SNPs (2 REC, 1 ADD, 1 MUL and 2 DOM). Only the top 5000 SNPs are
selected. The results are based on 200 replicates: MAF $q=0.2$ at the
functional locus when $D'=0.8$. Samples sizes are $r=s=1000$ for
GRR=1.25 and $r=s=500$ for GRR=1.5}
\begin{tabular*}{\textwidth}{@{\extracolsep{\fill}}lcd{3.1}ccccc@{}}
\hline
& &\multicolumn{3}{c}{$\bolds{D'=1.0}$}
&\multicolumn{3}{c}{$\bolds{D'=0.8}$}\\
\ccline{3-5,6-8}\\[-4pt]
\multicolumn{1}{@{}l}{\multirow{2}{20pt}[5pt]{\textbf{GRR} $\bolds{\lambda_2}$}}
&\multicolumn{1}{c}{\multirow{2}{30pt}[5pt]{\centering \textbf{Robust tests}}} &\multicolumn{1}{c}{\multirow{1}{40pt}[-5pt]{\centering \textbf{Prob}}}&\multicolumn{1}{c}{\multirow{2}{50pt}[6pt]{\centering \textbf{Ave. no. of true SNPs}}}&
\multicolumn{1}{c}{\multirow{2}{40pt}[6pt]{\centering \textbf{Mean of min ranks}}}
&\multicolumn{1}{c}{\multirow{1}{40pt}[-5pt]{\centering \textbf{Prob}}}&\multicolumn{1}{c}{\multirow{2}{50pt}[6pt]{\centering \textbf{Ave. no. of true SNPs}}}&
\multicolumn{1}{c@{}}{\multirow{2}{41pt}[6pt]{\centering \textbf{Mean of min
ranks}}}\\[2pt]
\hline
1.25 &$Z_{1/2}$ & 92.0 & 1.79 & 971 &58.5 &1.29 &1625\\
&GMS & 94.5 & 1.90 & 838  &56.0 &1.29 &1488\\
&$\mbox{MAX}_3$ & 90.5 & 1.80 & 909  &48.0 &1.28 &1435\\
&MIN2 & 89.5 & 1.79 & 934  &51.0 &1.25 &1550\\
&$T_{\chi^2}$ & 86.5 & 1.69 & 960  &46.5 &1.22 &1680\\[5pt]
1.50 &$Z_{1/2}$ & 99.5 & 2.71 & 186  & 83.0& 1.49&1041\\
&GMS &100.0 & 2.99 & 178  & 85.0& 1.54&1111\\
&$\mbox{MAX}_3$ & 99.5 & 2.83 & 205  & 80.0& 1.48&1183\\
&MIN2 &100.0 & 2.78 & 234  & 80.0& 1.50&1113\\
&$T_{\chi^2}$ &100.0 & 2.71 & 286  & 75.0& 1.46&1244\\
\hline
\end{tabular*}
\end{table*}

\begin{table*}[b]
\caption{\label{ranks2}Genome-wide scans of 300,000 SNPs containing 6
true SNPs (1 REC, 2 ADD, 2 MUL and 1 DOM). Only the top 5000 SNPs are
selected. The results are based on 200 replicates: MAF $q=0.2$ at the
functional locus and $D'=0.8$. Samples sizes are $r=s=1000$ for
GRR=1.25 and $r=s=500$ for GRR=1.5}
\begin{tabular*}{\textwidth}{@{\extracolsep{\fill}}lcccd{4.0}ccc@{}}
\hline
&&\multicolumn{3}{c}{$\bolds{D'=1.0}$}
&\multicolumn{3}{c}{$\bolds{D'=0.8}$}\\
\ccline{3-5,6-8}\\[-4pt]
\multicolumn{1}{@{}l}{\multirow{2}{20pt}[5pt]{\textbf{GRR} $\bolds{\lambda_2}$}}
&\multicolumn{1}{c}{\multirow{2}{30pt}[5pt]{\centering \textbf{Robust tests}}} &\multicolumn{1}{c}{\multirow{1}{40pt}[-5pt]{\centering \textbf{Prob}}}&\multicolumn{1}{c}{\multirow{2}{50pt}[6pt]{\centering \textbf{Ave. no. of true SNPs}}}&
\multicolumn{1}{c}{\multirow{2}{40pt}[6pt]{\centering \textbf{Mean of min ranks}}}
&\multicolumn{1}{c}{\multirow{1}{40pt}[-5pt]{\centering \textbf{Prob}}}&\multicolumn{1}{c}{\multirow{2}{50pt}[6pt]{\centering \textbf{Ave. no. of true SNPs}}}&
\multicolumn{1}{c@{}}{\multirow{2}{41pt}[6pt]{\centering \textbf{Mean of min
ranks}}}\\[2pt]
\hline
1.25 &$Z_{1/2}$ & 88.0 & 1.72 & 897  & 49.5& 1.31&1564\\
&GMS & 87.0 & 1.79 & 797  & 53.5& 1.27&1630\\
&$\mbox{MAX}_3$ & 82.5 & 1.64 & 846  & 47.0& 1.24&1702\\
&MIN2 & 86.0 & 1.66 & 932  & 48.5& 1.25&1899\\
&$T_{\chi^2}$ & 83.0 & 1.50 &1030  & 41.5& 1.20&1847\\[5pt]
1.50 &$Z_{1/2}$ & 99.0 & 2.46 & 349 & 76.5& 1.48&1083\\
&GMS & 99.5 & 2.61 & 355  & 76.0& 1.47&1005\\
&$\mbox{MAX}_3$ & 98.0 & 2.34 & 379 & 73.0& 1.40&1103\\
&MIN2 & 99.5 & 2.35 & 434  & 74.0& 1.38&1105\\
&$T_{\chi^2}$ & 97.0 & 2.21 & 485  & 66.5& 1.31&1179\\
\hline
\end{tabular*}
\end{table*}

\begin{table*}
\caption{Ranks of SNPs with strong association of seven
diseases in WTCCC (\protect\citeyear{WTCCC}), Table 3}\label{strong}
\begin{tabular*}{\textwidth}{@{\extracolsep{\fill}}lcd{2.0}d{3.0}d{3.0}d{3.0}d{3.0}d{3.0}@{}}
\hline
\textbf{Disease}& \textbf{SNP ID} & \multicolumn{1}{c}{\textbf{chrom}} & \multicolumn{1}{c}{$\bolds{Z_{1/2}}$} & \multicolumn{1}{c}{$\bolds{T_{\chi^2}}$}
& \multicolumn{1}{c}{\textbf{GMS}} & \multicolumn{1}{c}{$\mathbf{MAX_3}$} & \multicolumn{1}{c@{}}{\textbf{MIN2}}\\
\hline
BD &rs420259 &16 & 269 &22 &19 &20 &23\\[5pt]
CAD &rs1333049 &9 & 9 &25 &24 &24 &25\\[5pt]
CD &rs11805303 &1 & 14 &28 &23 &24 &24\\
&rs10210302 &2 & 6 &15 &15 &16 &15\\
&rs9858542 &3 & 102 &58 &58 &61 &75\\
&rs17234657 &5 & 11 &25 &19 &20 &21\\
&rs1000113 &5 & 72 &92 &78 &82 &84\\
&rs10761659 &10 & 89 &115 &100 &107 &101\\
&rs10883365 &10 & 50 &65 &59 &62 &61\\
&rs17221417 &16 & 25 &37 &35 &37 &38\\
&rs2542151 &18 & 69 &84 &77 &80 &81\\[5pt]
RA &rs6679677 &1 & 50 &72 &71 &69 &70\\
&rs6457617 &6 & 5 &13 &8 &8 &13\\[5pt]
T1D &rs6679677 &1 & 129 &137 &133 &136 &135\\
&rs9272346 &6 & 3 &6 &5 &5 &5\\
&rs11171739 &12 & 339 &361 &342 &357 &354\\
&rs17696736 &12 & 233 &245 &238 &243 &242\\
&rs12708716 &16 & 521 &534 &517 &534 &530\\[5pt]
T2D &rs9465871 &6 & 31 &41 &49 &44 &45\\
&rs4506565 &10 & 10 &17 &17 &17 &16\\
&rs9939609 &16 & 24 &38 &36 &36 &37\\
\hline
\end{tabular*}
\end{table*}

\section{Applications to WTCCC Data}\label{sec6}

We apply the five robust tests to a genome-wide scan using more than
300,000 SNPs after quality control. The study was originally conducted
by\break WTCCC (\citeyear{WTCCC}) for seven diseases (type 1 diabetes---T1D, type 2
diabetes---T2D, coronary heart disease---CHD, hypertension---HT,
bipolar disorder---BD,\break rheumatoid arthritis---RA and Crohn's disease---CD).
About 2000 cases were used for each disease and 3000 controls
were shared for the seven diseases.
WTCCC (\citeyear{WTCCC}) used MIN2 to test for association after the quality
control. They obtained two tables presenting SNPs with strong
associations with\break  $\mathrm{MIN2}<5\times10^{-7}$ (Table 3 of WTCCC,
\citeyear{WTCCC}) and SNPs with moderate associations with $5\times10^{-7} \le
\mathrm{MIN2}<5\times10^{-5}$ (Table 4 of WTCCC, \citeyear{WTCCC}). We reanalyze
these data by ranking all SNPs after our quality control. The goal of
this application is to demonstrate the efficiency robustness of
different test statistics, not to find SNPs with associations that were
not reported in WTCCC (\citeyear{WTCCC}).

In our application, for each of the seven diseases, we rank all SNPs
after quality control (398,092 SNPs) using the five robust tests and
report the ranks of the SNPs that were reported to have strong
associations in WTCCC (\citeyear{WTCCC}), Table 3. Note that we do not know $D'$ in
reality, nor do we know the number of functional loci and their modes
of inheritance. Our results are reported in Table \ref{strong}. The
results show that SNPs with strong associations are all ranked on the
top 5000 SNPs. The CATT is least robust among the five robust tests as
shown by the rank 269 for BD, while the ranks by the other methods are
less than 25. The GMS tends to have smaller ranks than $\mathrm
{MAX}_3$, and MIN2 tends to have ranks between the CATT and Pearson's
test, which often have higher ranks than the GMS.

We also studied the ranks of SNPs with moderate associations reported
in WTCCC (\citeyear{WTCCC}), Table 4. The detailed results are not shown here, but
summarized below. Similar patterns are also observed, although, for
several SNPs, the CATT has large ranks. For example, for BD, the CATT
has rank 147,769 for SNP rs6458307 on chromosome 6, while the ranks of
other tests for this SNP are less than 150. For T2D, the CATT has rank
197,064 for SNP rs358806 on chromosome 3, while the other tests have
ranks less than 100. All ranks of SNPs with either strong or moderate
associations are less than 5000, and only one SNP (rs17166496 for T1D
on chromosome 5) is ranked more than 5000 by $\mathrm{MAX}_3$ and the
GMS. The actual ranks for this SNP are 5521 for the GMS and 6063 for
$\mathrm{MAX}_3$, 652 for Pearson's test, 724 for MIN2, but 245,454 for
the CATT. The underlying genetic model for this SNP could be outside of
the constrained genetic model that we considered here, for example,
overdominant or underdominant for which it is known that Pearson's test
is robust (Zheng, Joo and Yang, \citeyear{zheng09}; Joo et al., \citeyear{joo}). In addition, we
found that for those SNPs with small ranks based on Pearson's test, a
large rank using the CATT is always accompanied by a large value of the
HWDTT. This is due to the orthogonal decomposition of Pearson's test to
the HWDTT and $Z_{1/2}^2$ (Zheng et al., \citeyear{zheng08}). It is also interesting
to note that, even if a SNP has a rank smaller than those SNPs listed
in Table \ref{strong}, it does not mean the SNP has a true association
with a disease. That is, in GWAS, a SNP with smaller $p$-value does not
necessarily mean it has stronger association. In fact, many of these
SNPs with smaller ranks have not been confirmed to have true
associations (WTCCC, \citeyear{WTCCC}). This is because a very small number of SNPs
($<$100 SNPs) are associated with a disease in GWAS compared to the
number of null SNPs (more than 300,000 SNPs). Therefore, the
probability that test statistics of some null SNPs are greater than
those of all the associated SNPs is high (Zaykin and Zhivotovsky, \citeyear{zaykin05}).

\section{Discussion}

We studied some robust tests for case-control genetic association
studies. This approach stems from the classical robust procedures
studied in the 1970s which focused on the estimation of the location
parameter of a symmetric distribution. For a given family of underlying
distributions (or, here, genetic models), an estimate with a high (low)
minimum correlation, say, $>$0.80 ($<$0.50) with the optimal procedure,
indicates a greater (smaller) efficiency robustness. In early work, the
underlying distribution was assumed to range from the normal
distribution to the Cauchy distribution (Tukey, \citeyear{tukey} and Andrews et
al., \citeyear{andrews}). For this family of $t$-distributions, the robust estimate
of the location parameter was considered, because within the family of
distributions considered, it had minimum correlation with the optimal
procedure of about 0.60 (Gastwirth,\break \citeyear{gast1}). In case-control genetic
association studies, when the true genetic model is unknown and ranges
from the REC to the DOM models, the minimum correlation of any two
CATTs is about 0.30 (Freidlin et al., \citeyear{freidlin02}). This indicates that using
a single CATT for association is not robust, and tests that are robust
across a family of plausible genetic models are preferred.

Previous studies of robustness properties of test statistics for the
analysis of case-control genetic association studies have been focused
on the perfect (or complete) LD model, that is, the genetic marker
(SNP) is also the functional locus. In this article we studied genetic
models under a general imperfect (or incomplete) LD model with linkage
disequilibrium between linked marker locus and functional locus. The
perfect LD model is a special case. Under the imperfect LD model, we
found that a genetic model defined by the genotype relative risks at
the functional locus usually no longer remains the same genetic model
at the marker locus, except for the additive or multiplicative models.
The genetic model space at the marker locus is a subset of that at the
functional locus, resulting in smaller genotype relative risks at the
marker than at the functional locus. The power to detect a true
association is reduced when the linkage disequilibrium decreases, while
the model uncertainty increases, complicating the choice of a single
association statistic. Robust tests are shown to perform optimally in
this situation.

We also review some common efficiency robust tests for case-control
genetic associations and their usage in genome-wide scans. In
genome-wide scans, all SNPs are ranked by a test statistic or its
$p$-value (if the $p$-value is readily obtained) and the top-ranked SNPs
are selected for further analyses. Alternatively, as in WTCCC (\citeyear{WTCCC}),
some genome-wide threshold levels can be also used to select SNPs.
Multiple testing is an important issue in GWAS not only because one
tests 300,000 up to a million SNPs, but also because multiple tests are
available for each SNP (and there is no uniform most powerful test in
GWAS). Correcting for multiple testing remains challenging in the
analysis of GWAS (Roeder and Wasserman, \citeyear{roeder}), and the need for
independent replication studies (Kraft, Zeggini and Ioannidis, \citeyear{kraft09}) and proper
meta-analysis (Pfeiffer, Gail and Pee, \citeyear{ruth}) cannot be
overemphasized.\looseness=1

\section*{Acknowledgments}
The research of Dmitri Zaykin was supported in part by the Intramural
Research Program
of the NIH, National Institute of Environmental Health Sciences. We
would like to thank three reviewers for their helpful suggestions and comments.

\end{document}